# Crisis, Country, and Party Lines:

# Politicians' Misinformation Behavior and Public Engagement

Jingyuan Yu[*1], Emese Domahidi[1], Duccio Gamannossi degl'Innocenti[2], and Fabiana Zollo[3,4]

[1] Department of Economic Sciences and Media, Ilmenau University of Technology, Germany

[2] Department of Economics and Management, University of Padua, Italy

[3] Department of Environmental Sciences, Informatics and Statistics, Ca' Foscari University of Venice, Italy

[4] The New Institute Center for Environmental Humanities, Italy

## Author Note

Jingyuan Yu  https://orcid.org/0000-0002-9400-4859

Emese Domahidi  https://orcid.org/0000-0003-3530-797X

Duccio Gamannossi degl'Innocenti  https://orcid.org/0000-0002-3653-216X

Fabiana Zollo  https://orcid.org/0000-0002-0833-5388

The research reported in this article was funded by Deutsche Forschungsgemeinschaft (DFG, German Research Foundation) under project number 458225198.

Correspondence concerning this article should be addressed to Jingyuan Yu, Ehrenbergstraße 29, 98693 Ilmenau, Germany. Email: jingyuan.yu@tu-ilmenau.de

**Abstract**

Politicians with large media visibility and social media audiences have a significant influence on public discourse. Consequently, their dissemination of misinformation can have profound implications for society. This study investigated the misinformation-sharing behavior of 3,277 politicians and associated public engagement by using data from X (formerly Twitter) during 2020–2021. The analysis was grounded in a novel and comprehensive dataset including over 400,000 tweets covering multiple levels of governance—national executive, national legislative, and regional executive—in Germany, Italy, the UK, and the USA, representing distinct clusters of misinformation resilience. Striking cross-country differences in misinformation-sharing behavior and public engagement were observed. Politicians in Italy (4.9%) and the USA (2.2%) exhibited the highest rates of misinformation sharing, primarily among far-right and conservative legislators. Public engagement with misinformation also varied significantly. In the USA, misinformation attracted over 2.5 times the engagement of reliable information. In Italy, engagement levels were similar across content types. Italy is unique in crisis-related misinformation, particularly regarding COVID-19, which surpassed general misinformation in both prevalence and audience engagement. These insights underscore the critical roles of political affiliation, governance level, and crisis contexts in shaping the dynamics of misinformation. The study expands the literature by providing a cross-national, multi-level perspective, shedding light on how political actors influence the proliferation of misinformation during crisis.

**Main**

Social media platforms are a primary source of news and information[1]. Through either active searches or unintentional exposure, news shared by political actors, such as prominent politicians and government agencies, has become an important part of citizens' media intake[2]. In this hybrid media system, traditional (e.g., newspapers) and digital media (e.g., social media) coexist and interact, while the boundaries between the two types of media are increasingly blurred[3]. The role of professional journalists as primary newsmakers has been challenged by the emergence of social media[3], which enabled politicians and others to join the news-making ecosystem as new actors[4].

Among the types of information shared by politicians, misinformation, broadly defined as any information that is misleading, unreliable, and/or turns out to be false[5], has received considerable attention[5,6]. Although misinformation represents only a small fraction of the messages shared by politicians[6,7], its societal impact remains substantial, as it may contribute to the spread of science denial[8] or the promotion of extreme political ideologies[9].

Due to variations in political structures, media traditions, and economic development in different regions and countries, the hybrid media system has distinct cross-territorial characteristics[10,11]. These distinctions have informed theorization of three clusters of countries, each exhibiting varying levels of resilience to misinformation[11]. The media-supportive cluster includes Western European democracies (e.g., Germany and the United Kingdom [UK]) with low levels of polarization and populist communication along with high levels of trust in established news media. This cluster is the most resilient to online misinformation[11]. The polarized cluster, consisting of Southern European countries such as Italy, is characterized by relatively high levels of political polarization, prevalent populist communication, and low trust in established news media. These factors make it more susceptible to online misinformation[11]. The United States of America (USA) represents a distinct cluster, marked by a politicized and fragmented media environment with similarly low trust in traditional media. This cluster is considered the most vulnerable to online misinformation due to high levels of polarized and populist communication and diminishing trust in established news sources. The size of the U.S. media market also incentivizes the production and dissemination of attention-triggering content[11].

Although countries exhibit varying levels of resilience to misinformation, the spreading of online misinformation is widely acknowledged as a global issue[12] and poses significant threats during crises. Even countries with high levels of misinformation resilience can be vulnerable to its influence[13]. Tackling misinformation at the global level requires comparative research to assess its impact and to develop effective mitigation strategies.

Empirical evidence on the spread of misinformation by politicians and its consequences remains inconclusive[4,14], and research beyond the United States is notably limited[15]. Our study therefore examined the misinformation-sharing behavior of politicians in Germany, Italy, the UK, and the USA. These countries were selected to represent distinct clusters of misinformation resilience[11] and diverse geopolitical contexts, including both EU member states (Germany, Italy) and non-EU countries (UK, USA). EU member states have comparatively stricter regulations on online misinformation[16].

Political authority is multi-layered in most countries, with decision-making power distributed across executive and legislative branches at national and regional levels. This highlights another critical research gap, as existing studies have tended to focus on national legislative politicians[7,14], often overlooking the roles of executive and regional politicians. Nonetheless, these groups can play a substantial role in the spread of misinformation, with potentially far-reaching consequences. For instance, executive politicians typically have greater media visibility and a larger social media audience than legislative politicians[17], which may amplify the reach of misinformation they share. Similarly, regional politicians' communication on social media can strongly shape public opinion on local issues[18] and make misinformation they share particularly impactful in their areas.

To address these overlooked dimensions, we analyzed and compared the misinformation-sharing behavior of national executive (NE), national legislative (NL), and regional executive (RE) politicians in the selected countries. The goal was to provide a more comprehensive understanding of how misinformation proliferates across different political roles and levels of governance. Our study was focused on the years 2020 and 2021, a period marked by significant political and societal events, including elections, government transitions, and the COVID-19 pandemic, which raised concerns about misinformation and the so-called 'infodemic'[19,20]. We selected X (formerly Twitter) for the research due to its extensive use by politicians in the selected countries[1]. Building on the two identified research gaps, our study addressed the following research question:

- **RQ1:** *How much (mis)information was shared by the politicians a) across the four selected countries and b) at the three levels of political hierarchy?*

Politicians may prioritize advancing their agendas over ensuring factual accuracy, potentially using misinformation as a strategic tool to achieve political goals[4,21,22]. Due to their high visibility, misinformation shared by politicians might attract substantial public engagement[23], which can in turn increase the perceived credibility of the content[24,25]. This dissemination could be driven more by the authority associated with politicians than by the factual accuracy of the information[23,26]. However, it remains unclear whether misinformation shared by politicians consistently outperforms reliable information in engagement.

- **RQ2:** *To what extent did the shared (mis)information attract public engagement in the countries?*

Existing research suggests that misinformation dynamics are context-dependent[13], often shaped by major political and crisis events such as elections or natural disasters[27].

Unlike political events, which tend to have predictable patterns (e.g., elections) or antecedents (e.g., Brexit), crises typically occur abruptly, creating an information vacuum. During such times, heightened emotions and unmet informational needs can create fertile ground for misinformation to thrive[28]. This was particularly evident during the COVID-19 pandemic, where the overabundance of information, including misinformation, was described as an infodemic[19,29]. However, it remains uncertain to what extent crises amplify the dissemination of misinformation by politicians and whether misinformation elicits greater public engagement compared to other types of content.

- **RQ3**: *To what extent was crisis-related misinformation a) shared and b) engaged with, compared to other types of misinformation, in the four countries?*

Studies have indicated that conservative[14] and far-right politicians[7] are the most frequent sources of misinformation, especially during periods of heightened political or societal tensions[13,30]. The societal and political turbulence of 2020–2021 likely shaped how politicians from different parties communicated with the public[31], and the extent to which they engaged in sharing misinformation[7,31]. This context underscores the need to reexamine previous research findings and evaluate how these dynamics impacted public engagement during this period.

- **RQ4:** *To what extent did different political parties share misinformation on X during 2020 and 2021, and how much public engagement did this misinformation attract?*

Addressing these four questions will provide insights into the overall landscape of the misinformation-sharing behavior of politicians in various countries and political hierarchies over two years of severe crises. By analyzing public engagement, we have identified interaction patterns that contribute to the dynamics of misinformation dissemination. Our study thus offers comparative insights into how misinformation propagates in diverse contexts and thus enhances understanding of the roles of political hierarchy and national context in the sharing of misinformation and its impact on public discourse.

## Results

To investigate politicians' misinformation-sharing behavior across the four selected countries and the three theorized clusters of misinformation resilience, we start by calculating the proportion of misinformation shared by politicians (Fig. 1, top panel). To distinguish between misinformation and reliable information, we used NewsGuard's trust scores. A source is classified as misinformation if its trust score is below 60 and as reliable if the score is 60 or higher. Each politician's tweet was categorized based on the trust score of the source it referenced (see Methods for details). Among the countries analyzed, Italy exhibited the highest average share of misinformation ($M = 0.049$, $SD = 0.147$), approximately 2.2 times higher than the USA ($M = 0.022$, $SD = 0.056$), 4.5 times higher than Germany ($M = 0.011$, $SD = 0.055$), and 12.3 times higher than the UK ($M = 0.004$, $SD = 0.036$). An ANOVA test revealed a significant main effect of country on misinformation-sharing behavior ($F(3, 3116) = 42.65$, $p < .001$). Post hoc comparisons using the Tukey HSD test indicated that Italy exhibited a significantly higher proportion of misinformation compared to the other countries (detailed results can be found in the supplementary information [SI]). Further analysis on the three levels of political hierarchy (Fig. 1, bottom panel) revealed that NL politicians shared the most misinformation. Among RE politicians, the highest proportion of misinformation was observed in the USA (11.1%), followed by the UK (6.8%), with notably lower levels in Germany and Italy. Pairwise comparisons using Tukey HSD tests indicated no statistically significant differences in misinformation-sharing behavior across the three levels of political hierarchy within Italy, the UK, or the USA. Germany was not considered in this analysis, as no misinformation was shared by either NE or RE politicians during the study period.

**Fig. 1: Misinformation-sharing behavior by country and levels of political hierarchy**

Top panel: Proportion of misinformation shared at the politician level in the four countries. Bottom panel: Distribution of shared (mis)information among national executive, national legislative, and regional executive politicians in the four countries.

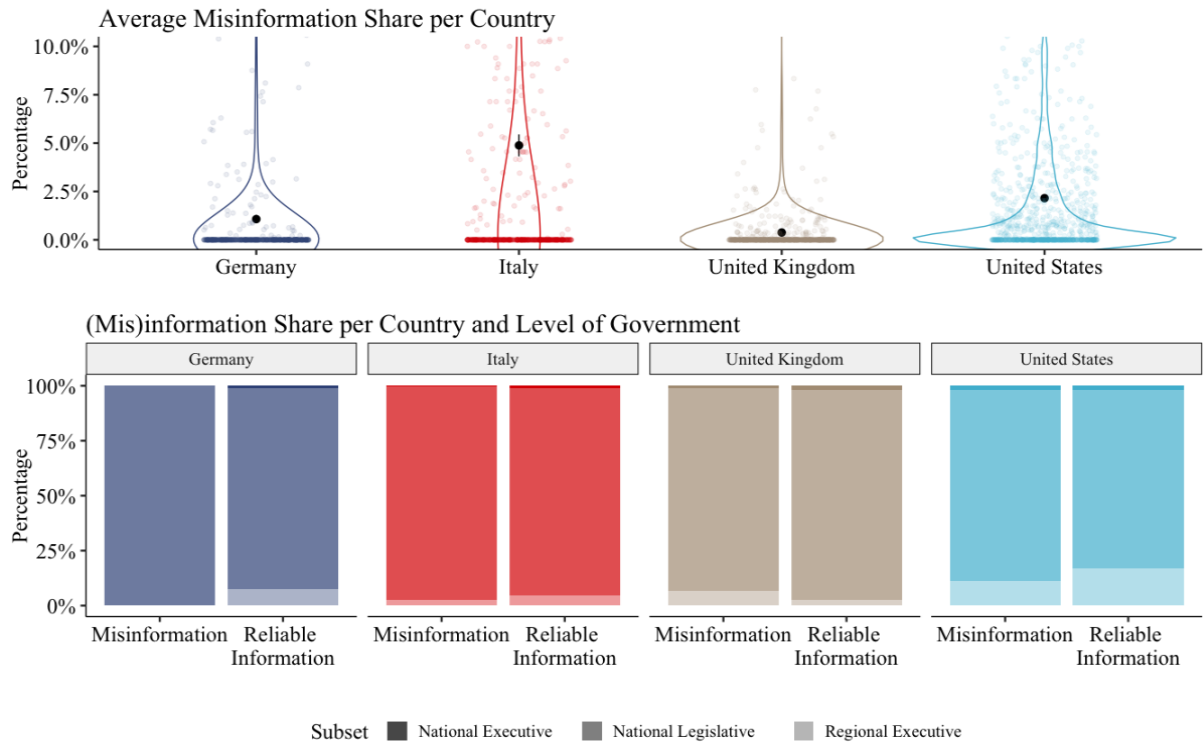

The measure of public engagement used in our analysis is the sum of retweets and likes for each tweet. Retweets are generally interpreted as endorsements[32], while likes indicate general user interest[33]. Fig. 2 compares user engagement with reliable information versus misinformation over time, offering insights into patterns and variations across countries. Fig. 2a) illustrates the cumulative sum of engagement of the two types of information over time, along with a linear trend. The slope of the trend line, represented by the beta coefficient, reflects the ratio of engagement between the two types of information, with a steeper slope indicating higher engagement with misinformation. Deviations from the trend line highlight tweets that went viral with spikes in public engagement.

For the four countries, the beta coefficient consistently falls below 1, indicating that reliable information attracted higher engagement than misinformation overall. However, statistically significant cross-country variations existed, as confirmed by pairwise z-tests across beta values (all significant at the 1% level). The smallest difference was observed between Italy and the USA (z-statistic = 55.230; detailed results are provided in the SI). Both countries exhibited high engagement with misinformation: The USA shows a beta of 4.3%, while Italy registers a considerably higher beta of 9.6%, representing a 2.2-, 12-, and 48-fold increase over the US, German, and UK figures, respectively.

In contrast, the results for the UK and Germany, with beta values of 0.2% and 0.8%, respectively, indicate that misinformation received only a fraction of the engagement that reliable information did. Since these comparisons summarize both the volume of tweets and the average engagement per tweet, we investigated the latter aspect (Fig. 2b) by comparing the cumulative average engagement over time. In Fig. 2b, beta coefficients are notably higher than in Fig. 2a. Statistically significant differences between countries persisted (1% level), with the closest comparison again between Germany and the UK (z-statistic = 62.695). The UK and Germany remained ranked similarly, with misinformation receiving 20.3% and 38.4% of the average cumulative engagement compared to reliable information. Italy and the USA, however, switched their positions: In Italy, misinformation posts received 1.09 times the average engagement of reliable information, while in the USA, misinformation drew 2.55 times the average engagement. These findings highlight significant cross-country variability in user engagement with (mis)information and align with the patterns suggested by the three clusters identified in the literature[11].

**Fig. 2: Public engagement with (mis)information per country**

Panel a): Cumulative sum of engagement with misinformation versus reliable information. The y-axis stands for misinformation, and x-axis represents reliable information. Statistical differences are significant across the studied countries. Italy had the highest and the UK the lowest sum of misinformation engagement.

Panel b): Cumulative average engagement with misinformation versus reliable information. The y-axis and x-axis represent misinformation and reliable information, respectively. Statistical differences are significant across the studied countries. The USA had the highest and the UK the lowest average misinformation engagement.

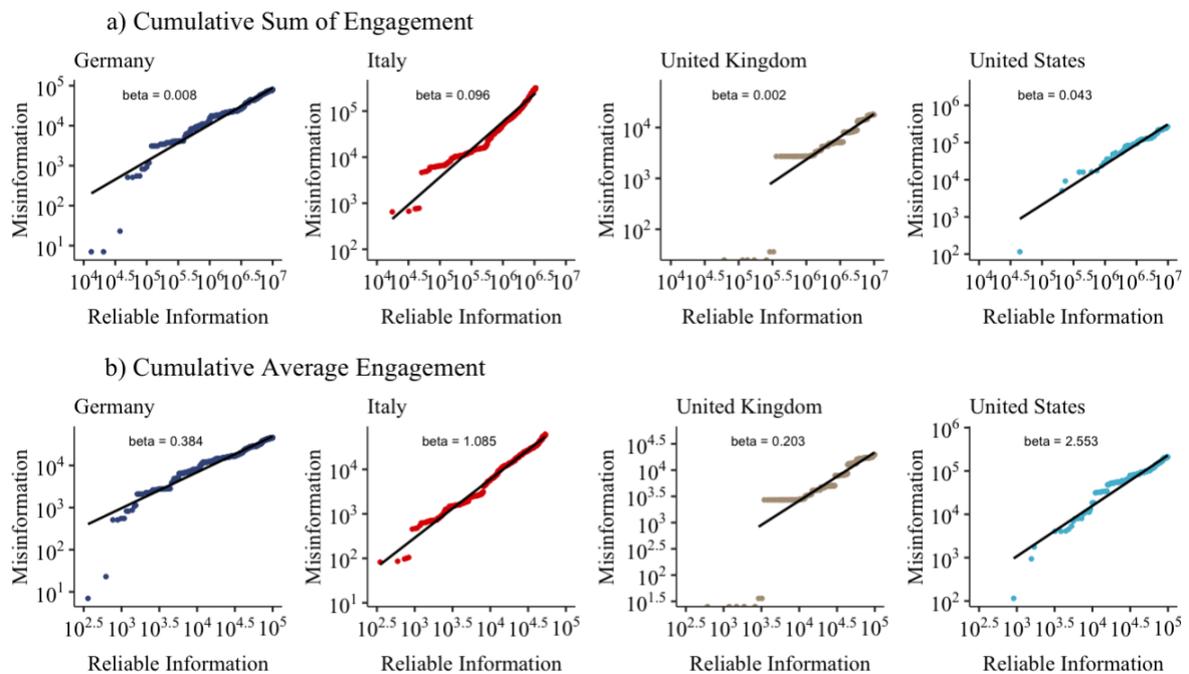

We next distinguished between crisis-related and general misinformation shared by politicians. The classification of the two categories was based on the official annotation from X and enhanced by a set of manually validated keyword lists consisting of relevant uni-, bi-, and trigrams (see Methods for details). Fig. 3a) shows the relative proportions of these two categories of misinformation shared by politicians in each country. For each politician, we calculated the relative share of crisis-related and general misinformation within their total misinformation output. A two-way ANOVA revealed significant effects for both country ($F(3, 1217) = 115.47$, $p < .001$) and misinformation type ($F(1, 1217) = 21.83$, $p < .001$). The interaction between country and misinformation type is also statistically significant ($F(3, 1217) = 5.46$, $p < .001$). Post hoc analysis using the Tukey HSD test indicated that Italian politicians shared significantly more crisis-related misinformation compared to general misinformation ($p < .001$). However, no significant differences between misinformation types were observed in the other three countries (further details are provided in the SI).

Fig. 3b and 3c compare the cumulative sum and cumulative average engagement with crisis-related and general misinformation over time. The y-axis represents the cumulative engagement for crisis-related misinformation, while the x-axis represents general misinformation. In the four countries, the beta coefficient consistently falls below 1 in both panels, indicating that from both perspectives, crisis-related misinformation attracted less engagement than general misinformation. However, variations among the countries existed. From the cumulative sum perspective (Fig. 3b), Germany had the highest engagement with crisis-related misinformation, with a beta coefficient of 41.2%, which is 3.6 times more than that of the USA—the country with the lowest cumulative sum engagement. Regarding cumulative averages, Italy had the highest average engagement with crisis-related misinformation, with a beta coefficient of 79%, which is 2.9 times more than that of the USA, which is still lowest among the countries. Statistical analyses of cross-country variations showed that in both aspects, all the pairs exhibit significant differences (for more details, refer to SI).

**Fig. 3: Comparison of crisis-related and general misinformation by country**

Panel a): Proportions of crisis-related versus general misinformation shared by politicians in four countries. Statistical analyses reveal significant cross-country variations, with Italian politicians sharing significantly more crisis-related misinformation than general misinformation.

Panel b): Cumulative sum of engagement with crisis-related versus general misinformation over time. The y-axis represents engagement with crisis-related misinformation, while the x-axis represents engagement with general misinformation. Significant differences in engagement were observed, with Germany showing the highest and the USA the lowest engagement with crisis-related misinformation.

Panel c): Cumulative average engagement with crisis-related versus general misinformation over time. The x-axis and y-axis represent general and crisis-related misinformation, respectively. Significant differences in engagement were observed, with Italy showing the highest and the USA the lowest engagement with crisis-related misinformation.

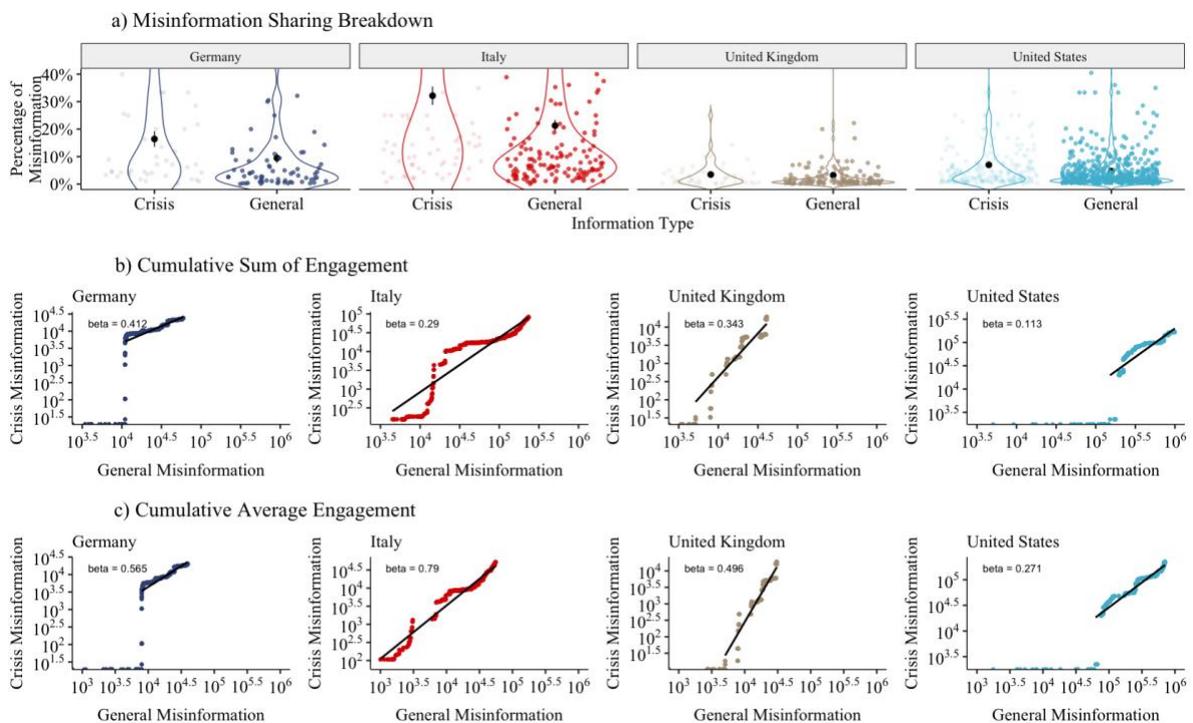

We then compared the misinformation sharing of political parties. Fig. 4 presents the share of misinformation attributed to political parties in the countries, along with the distribution of public engagement for each party. We performed this analysis on NL politicians, focusing on the parties responsible for more than 90% of misinformation tweets within each country. Independent politicians and politicians who did not clearly belong to any party (e.g., Fraktionslos politicians in Germany) were excluded. The stacked bar plots illustrate the overall share of misinformation for these dominant parties in each country. Notably, far-right (Alternative für Deutschland [AfD, Germany], Fratelli d'Italia [FdI, Italy]) and conservative (Republican [USA], Conservative [UK]) parties were responsible for the largest shares of misinformation. AfD accounted for 95.5% of misinformation shared in Germany, FdI for 70.7% in Italy, the Conservative Party for 57.3% in the UK, and the Republican Party for 76.4% in the USA. When comparing the parties within each country, AfD stood out in Germany, where 8.2% (n = 951) of the links shared by AfD politicians were misinformation—41 times higher than the second-largest misinformation-sharing group, Die Linke (n = 23, 0.2%) and 82 times higher than the third-largest group, CDU/CSU (n = 7, 0.1%). In Italy, FdI led with 41.7% (n = 2,747) of its shared links being misinformation, 4.5 times higher than the second-largest group, Lega (n = 927, 9.2%), and 13.9 times higher than the third-largest, Forza Italia (n = 82, 3%). In the UK, the Conservative Party (n = 149, 0.5%) shared five times more misinformation than the Labour Party (n = 59, 0.1%). In the USA, Republicans (n = 2,501, 4.6%) shared 4.6 times more misinformation than Democrats (n = 774, 1.0%). Consistent with previous research[7], these results show that (far-)right parties were significant contributors to misinformation across all the selected countries. However, a notable contrast emerged when comparing Italy to the other countries, particularly in the proportion of misinformation shared by FdI. Nearly 42% of FdI's content was deemed unreliable, a stark difference compared to the far-right parties in other countries whose posts barely reached the 10% mark.

In terms of public engagement, Fig. 4 shows the complementary cumulative distribution functions (CCDFs) of the engagement received by misinformation posts for each party and country. Most of these parties fit a heavy-tailed log-normal distribution (for details see SI) indicating that while most misinformation posts received low or moderate engagement, a small portion of them went viral. As mentioned, far-right and conservative parties dominated misinformation sharing. However, notable differences between political parties emerged, particularly in Italy and the USA. For Italy, Lega exhibited a slower decline in its CCDF compared to FdI, indicating that a larger proportion of Lega's misinformation posts garnered higher engagement. This contrasts with FdI, where there was a disconnect between the volume of misinformation shared and the relatively low engagement it generated. In the USA, Republicans also experienced higher engagement with their misinformation compared to Democrats, with a larger proportion of Republican misinformation posts receiving significant attention.

**Fig. 4: Misinformation share and engagement distribution by political party**

Left plot: Complementary cumulative distribution function of misinformation engagement per political party in each country.

Right plot: Stacked bar plot showing the proportion of misinformation share per political party in each country.

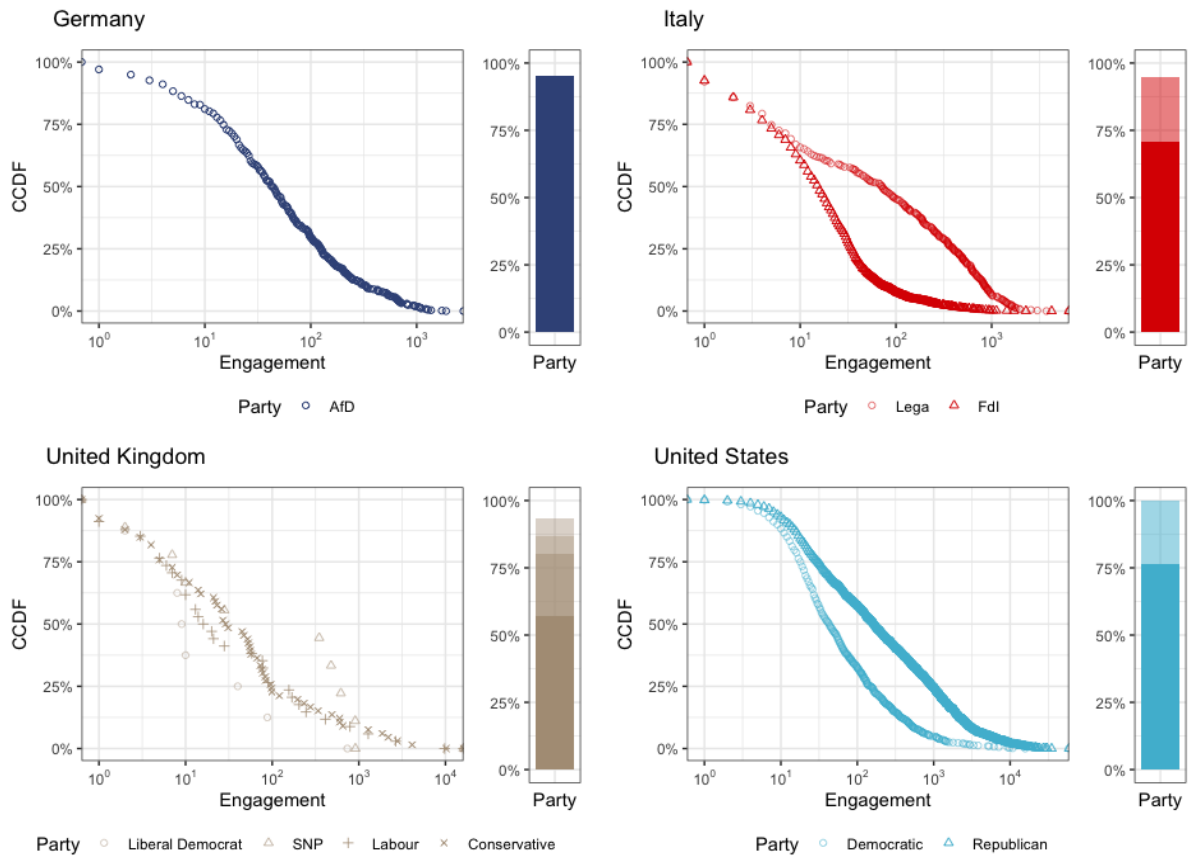

**Discussion and Conclusion**

Our findings show a differentiated picture of the role of politicians as sources of misinformation in four countries. To answer RQ1, politicians from Italy and the USA shared the most misinformation among the countries, with legislative politicians being the main forces of misinformation sharing. Executive politicians at both national and regional levels shared only a limited amount of misinformation, particularly in Germany and Italy. Addressing RQ2, misinformation generally received less engagement compared to reliable information, amounting to one-fifth and two-fifths of the engagement levels in the UK and Germany, respectively. In Italy, engagement levels were similar for both types of information, while in the US, engagement with misinformation was more than 2.5 times higher than with reliable information. However, due to the relatively low volume of misinformation, the cumulative engagement with misinformation remained lower than that of reliable information, peaking at 12% in Italy, a modest 3% in the USA, and being negligible in both Germany and the UK. These findings contradict previous evidence that the USA was a unique country where politicians shared much more misinformation than in other countries[7]. Instead, the findings align with research concluding that both Italy and the USA have low levels of trust in established news media, coupled with high levels of polarization and populist communication[11], which contribute to increase misinformation sharing and engagement.

Regarding RQ3, crisis-related misinformation and general misinformation were shared to a similar extent in most of the countries. Italy stood out as the only case where the sharing of crisis-related misinformation was significantly higher than the sharing of general misinformation. Even though crisis-related misinformation tended to receive less public engagement than general misinformation in all the countries, there were cross-country differences. Italy had the highest public engagement in crisis-related misinformation among the countries. The elevated engagement with crisis-related content may reflect public concern or heightened sensitivity during the COVID-19 crisis[34], as Italy was one of the first European countries to be largely impacted by the pandemic[35]. In all countries studied, approximately half of the misinformation shared was related to crises, supporting the argument that misinformation sharing is largely event-driven[13]. However, our results suggest a need for a more nuanced understanding of the 'infodemic' narrative during COVID-19[19,29]. While reliable information dominated social media communication by politicians, a significant portion of misinformation was unrelated to crises. Furthermore, crisis-related misinformation attracted less engagement compared to general misinformation, which challenges some prevailing assumptions about its impact[19,36].

Regarding RQ4, at the party level, consistent with previous research[7], our results show politicians from far-right, conservative, and populist parties shared the most misinformation. Moreover, the heavy-tailed log-normal distribution of public engagement in most parties means that only a small amount of misinformation reached a large audience. This disproportionality suggests that those countering misinformation should prioritize addressing highly engaged and deep-impact posts rather than attempt to detect all instances of misinformation. Given the rapid rise of extremism in Europe and the USA, for which misinformation is considered a key driving factor[37], effective and appropriate online content moderation strategies are needed to protect democratic values.

This study has some limitations that also suggest avenues for future research. First, we adopted a source-based approach to classify misinformation by evaluating the reliability of its source (i.e., news domains). This evaluation was carried out by an independent organization using consistent criteria for the analyzed countries, which made it particularly suitable for large-scale cross-country comparisons. However, this approach does not account for misinformation generated through original content shared directly by politicians. To date, content-based approaches to detecting misinformation in large datasets are unfeasible and often unreliable, particularly in comparative studies. Future research could benefit from a greater focus on reliably classifying the content of posts, potentially by using large language models. These tools offer the opportunity to examine the misinformation generated by politicians, its prevalence in different countries and events, and its relative prevalence on social media compared to other sources. Second, our study was focused on a single platform due to the prominence of X in shaping political discourse during the analyzed period[38]. However, this focus may limit understanding of the broader landscape of online political communication[39]. Given the differences in platform affordances[40] and user demographics[41], patterns of misinformation sharing and public engagement are likely to vary across platforms. Future comparative studies should incorporate additional social media platforms to expand the scope of our findings. Third, this research was limited to high-income countries, which hinders the generalizability of its results on a global scale[42]. The proliferation of politician misinformation is not limited to Western Europe and the USA. In certain cases[43], the challenge of misinformation is even more pronounced. Therefore, scaling up the research scope by including low- and middle-income countries is essential.

Based on an extensive dataset that accounts for a diverse range of countries with different political and media environments, as well as a comprehensive list of politicians at different levels of political hierarchies, this study systematically examined politicians' misinformation-sharing behavior and the corresponding public engagement during 2020 and 2021. Our findings provide a nuanced perspective on how factors such as political affiliation, governance level, and crisis context shape misinformation dynamics in various domains. Despite its limitations, our study provides a cross-national, multi-level analysis of misinformation dissemination and engagement patterns and sheds light on the role of political actors in shaping misinformation dissemination during crises.

## Methods

**Data collection and processing**

We deductively sampled NE, NL, and RE politicians from the four countries according to official websites and then identified their respective usernames on X. Our research data were subsequently retrieved using Twitter Academic API[44]. We included information about politicians who were newly elected or discontinued from their positions during 2020 and 2021 to build our list. Data that exceed this time range of position were excluded. In addition, accounts that had not posted in 2020 or 2021 were excluded.

For crisis-related posts, we focused on the COVID-19 pandemic. COVID-related messages were identified through automatic labeling and a complementary validated uni-, bi-, and trigram keyword list. We validated our results through manual coding, achieving recall rates of between 0.84 and 0.91 and precision rates between 0.83 and 0.93 in the four countries (details for the classification are available in SI).

**Misinformation classification**

We used both original tweets and retweets. We extracted domains from the attached URLs, decoding and expanding shortened URLs (e.g., bit.ly). Social media links (e.g., twitter.com, facebook.com) were removed. The processed domain lists were then matched against the NewsGuard rating database[45], a third-party independent resource for news domain credibility. According to NewsGuard, news media are evaluated based on nine key criteria related to credibility and transparency[45]. These criteria form the foundation for the rating system, which assigns a total score of 100. Domains with a score below 60 are considered misinformation.

We collected and processed a total of 1,771,518 URLs, of which 411,347 matched with the NewsGuard database as containing a link to an external news domain. These matched URLs served as the final data for our analyses. Table 1 reports the number of politician accounts and the number of (mis)information tweets in the four countries, with percentages in parentheses indicating each country's contribution to the total.

**Table 1**

*Number of Politician Accounts and (Mis)information Tweets by Country*

| Country | National Executive Politicians | National Legislative Politicians | Regional Executive Politicians | Total Politicians | Number of Tweets | Reliable information tweets | Misinformation tweets |
|---|---|---|---|---|---|---|---|
| Germany | 21 | 652 | 92 | 765 | 68,871 (16.74%) | 67,844 (16.95%) | 1,027 (9.28%) |
| Italy | 28 | 567 | 97 | 692 | 62,212 (15.12%) | 56,324 (14.07%) | 5,888 (53.23%) |
| United Kingdom | 38 | 815 | 50 | 903 | 116,354 (28.29%) | 115,974 (28.97%) | 380 (3.44%) |
| United States | 56 | 583 | 278 | 917 | 163,910 (39.85%) | 160,143 (40.01%) | 3,767 (34.05%) |
| Total | 143 | 2,617 | 517 | 3,277 | 411,347 | 400,285 | 11,062 |

Note: Politicians who had an inactive X account, i.e., haven't posted anything during 2020 and 2021, were excluded from our analysis